# Nanoscale Topographical Replication of Graphene Architecture by Artificial DNA nanostructures

Y. Moon,[1,a)] J. Shin,[2,a)] S. Seo,[1] J. Park,[1] S. R. Dugasani,[2] S. H. Woo,[3] T. Park,[1] S. H. Park,[1,2,b)] and J. R. Ahn[1,c)]

[1]*Department of Physics, Sungkyunkwan University, Suwon 440-746, Republic of Korea.*

[2] *Sungkyunkwan Advanced Institute of Nanotechnology (SAINT), Sungkyunkwan University, Suwon 440-746, Republic of Korea.*

[3] *College of Pharmacy, Chungnam National University, Daejeon 305-764, Republic of Korea.*

Despite many studies on how geometry can be used to control the electronic properties of graphene, certain limitations to fabrication of designed graphene nanostructures exist. Here, we demonstrate controlled topographical replication of graphene by artificial deoxyribonucleic acid (DNA) nanostructures. Owing to the high degree of geometrical freedom of DNA nanostructures, we controlled the nanoscale topography of graphene. The topography of graphene replicated from DNA nanostructures showed enhanced thermal stability and revealed an interesting negative temperature coefficient of sheet resistivity when underlying DNA nanostructures were denatured at high temperatures.

___________________________

[a)] Y. Moon and J. Shin contributed equally to the entire work.

[b)] Electronic mail: sunghapark@skku.edu.

[c)] Electronic mail: jrahn@skku.edu.

Graphene is a very fascinating material because of its unique mechanical and electronic properties.[1] One of the major challenges for graphene is controlling its electronic structure in a designed manner for various device applications. To control the electronic structure and unit size of graphene nanostructures, various approaches have been reported, including fabrication of graphene nanoribbons,[2,3] chemical functionalization of graphene,[4] strain applied to graphene by stretching or bending,[5-7] and nanoscale control of three-dimensional (3D) topography of graphene.[8,9] Among these, the 3D topographical control of graphene showed interesting phenomena such as a pseudo-magnetic field, which was observed in 3D strained graphene nanobubbles on Pt by scanning tunneling microscopy. The Dirac point energy gap of graphene was theoretically suggested to be controlled by periodic nanoscale ripples of graphene. The 3D topographical control of graphene at the nanoscale level is quite difficult because graphene intrinsically prefers a two-dimensional (2D) structure. Although graphene with local 3D topography was reported to exist in the form of nanobubbles on a Pt (111) surface,[8] ripples on chemical vapor deposited (CVD) graphene,[10-15] and corrugated structures on double strand deoxyribonucleic acids (ds-DNAs),[16] there are limits to the geometrical shapes that can be constructed.

It is not possible for graphene itself to produce the designed 3D structures except by creating artificial defects in graphene.[10,17] Designed templates with nanometer-scale precision are thus required to make various 3D graphene structures in a controlled manner. Deoxyribonucleic acid (DNA) nanotechnology has provided a platform to construct artificially designed nanostructures which were self-assembled with precisely controllable and programmable nanoscale features with the aid of oligonucleotide recognition.[18-27] Here, we demonstrate that the nanoscale 3D topography of graphene can be controlled in a designed manner by using artificially designed DNA nanostructures with a high degree of geometrical freedom. Two DNA nanostructures, a one-dimensional (1D) five-helix ribbon (5HR) structure[27] and a 2D double-crossover (DX) lattice,[22] were self-assembled in a solution during annealing. After formation of DNA nanostructures, the samples were deposited on a mica surface. For large scale topographical control, CVD graphene was adopted rather than mechanically exfoliated graphene. CVD graphene, which was grown on Cu foil, was transferred onto the DNA nanostructures on the mica surface; during this process, graphene nanostructures were replicated from DNA nanostructures (hereafter, replicated graphene-DNA nanostructures). After the production of the designed 3D topography of replicated graphene-DNA nanostructures, we further studied its thermal stability. The influence of temperature on its topography and electrical properties was verified by atomic force microscopy (AFM) and a four point probe, respectively.

Synthetic oligonucleotides were purchased from Integrated DNA Technologies (IDT, Coralville, IA), purified by high performance liquid chromatography (HPLC). The details can be found on www.idtdna.com. Complexes were formed by mixing a stoichiometric quantity of each strand in physiological buffer, 1×TAE/$Mg^{2+}$. Annealing of each DNA nanostructure were done by mixing the appropriate strands in an AXYGEN-tube and placing this in 1.5 L of boiled water in a Styrofoam box. The water was slowly cooled from 95 °C to 25 °C for at least 12 hours to facilitate hybridization. Final concentrations of DNA nanostructures were 200 nM. 5 μL of samples and 45 μL of 1×TAE/$Mg^{2+}$ buffer were placed on a freshly cleaved mica surface for 2 minutes and we blew it off by applying nitrogen gas. Graphene was prepared in a CVD chamber at 1000 °C on copper foil (25μm, Alfa Aesar). Copper foil was etched at 1000 °C in 100 sccm hydrogen atmosphere 2 hours. Subsequently, graphene was grown for 30 minutes at 1000 °C with 20 sccm methane and 100 sccm hydrogen atmospheres. The prepared graphene was transferred on a mica surface, which was covered with dried DNA nanostructures, by using poly(methyl methacrylate) (PMMA) (Sigma-Aldrich, Mw ~996). All AFM images were taken by Nanoscope (Veeco Inc.) in air intermittent contact mode and samples were fixed on a metal puck. A silver pasted four-probe technique was used to measure the sheet resistivity of graphene replicated by 5HRs through a Model 370 AC Resistance Bridge in a heating system from 30 °C up to 175 °C.

Fig. 1(a) shows a schematic representation of the replication of graphene by the DNA nanostructures (5HR). The light green cylinders on the mica surface indicate the DNA duplexes, which belong to a single 5HR nanostructure. The mesh shape on the top of the DNA nanostructures refers to graphene. The 5HR nanostructure is composed of six single strands, which are mutually linked to form a 1D five duplex ribbon structure. To understand the 3D topography of the transferred graphene onto the DNA nanostructures-covered mica surface, AFM images were acquired. All AFM images were acquired through tapping mode in air. Fig. 1(d) and (e) are the AFM images of 5HRs and replicated graphene-5HRs on mica surfaces, respectively. Contrary to the randomly attached 5HR on mica surface (Fig. 1(d)), replicated graphene-5HRs are arranged (Fig. 1(e)). After the wet transfer of graphene, 5HRs under the graphene might be rearranged through the drying process of deionized water. The bright lines in the inset in Fig. 1(e) correspond to the replicated graphene-5HRs. These bright lines had different widths of approximately 30 and 55 nm, as shown in the inset of Fig. 1(e). The width of the 5HR, as measured from the inset in Fig. 1(d), was 15.36 nm. It means that the bright lines in Fig. 1(e) have one (~29.5 nm) or two (~55.6 nm) underlying 5HRs. The AFM images suggest that graphene can replicate 5HRs. Slightly wider 5HRs are shown in Fig. 1(d) compared to that of the 5HR nanostructure itself due to the finite radius of the AFM tip apex, where five duplexes have a width of approximately 10 nm.[16] We tried to replicate DX lattices, the 2D DNA nanostructure, to test broad replication possibility of various DNA nanostructures. In contrast with 1D 5HRs, the 2D DX lattice is made up of double-crossover motifs with exchanges between strands of opposite polarity. Fig. 1(f) and (g) are the AFM images of 2D DX lattice and replicated graphene-DX lattices, respectively. The images show that graphene can replicate 2D DNA nanostructure. The height of replicated graphene-DX lattice is 2.89 nm. It is roughly two times higher than DX lattice itself, 1.6 nm. This height difference between the DNA nanostructure and the replicated graphene-DNA nanostructure was enhanced in case of 5HRs; the height of 5HR itself is 1.08 nm and the height of replicated graphene-5HRs is between 2.83 and 6.84 nm. The height differences suggest that strain in graphene depends on the DNA nanostructure. As a result, we could make higher corrugated graphene topography using 5HRs. Some of the residues in the AFM images in Fig. 1(d) and (e) are ascribed to PMMA, which was used for the transfer process of graphene.[28] In the typical transfer process, the PMMA residues can be further removed by thermal treatments after acetone treatments.[18,30] However, the thermal treatments cannot be applied to this replication process because the DNA nanostructures are easily denatured by heat so that the PMMA residues are inevitable in this replication process.

Fig. 1(b) and (c) show Raman spectra before and after the replication of graphene by DNA nanostructures on the mica surface, respectively. For 5HRs on the mica surface (Fig. 1(b)), we observed three Raman scattering peaks with Raman shifts of 1020, 1081, and 1120 cm$^{-1}$. After the transfer of graphene on 5HRs on the mica surface (Fig. 1(c)), two typical Raman scattering peaks of graphene, called G and 2D, were observed with the three Raman scattering peaks originating from the pristine 5HRs. The G peak, with a Raman shift of 1580 cm$^{-1}$, and the 2D peak, with a Raman shift of 2680 cm$^{-1}$, originate from the breathing modes of sp$^2$ carbon atoms and two phonons with opposing momentum in the highest optical branch near the K point of the Brillouin zone of graphene.[30] The Raman spectra suggest that graphene can be transferred onto the DNA nanostructures on the mica surface.

The DNA nanostructures are disassembled easily even at moderate temperatures near 60 °C. The low thermal stability of DNA nanostructures has significantly limited their physical, chemical, and biological applications.[27, 31] Thus, the thermal stability of the graphene-covered DNA nanostructures was further studied systematically to examine the feasibility of graphene as a thermal protection atomic layer for DNA nanostructures. Fig. 2(a)-(c) show changes in the AFM images of pristine 5HRs on a mica surface with increasing temperature. The 5HRs started to be denatured and deformed after 5 minutes at a temperature of 60 °C and were completely deformed after 30 minutes at the same temperature. The low thermal stability is due to the weak hydrogen bonds, with a binding energy of approximately 5 kJ/mole between DNA base pairs.[27, 31, 32] Fig. 2(d)-(f) show the changes in the AFM images of replicated graphene-5HRs with increasing temperature. In comparison to the pristine 5HRs, the topography of the replicated graphene-5HRs persisted even after 30 minutes at a temperature of 60 °C, at which point DNA nanostructures without graphene are completely denatured. When the temperature was raised to 150 °C, the 3D topography of the replicated graphene was mostly flattened. The comparative studies of the 5HRs with and without graphene suggest that the replicated graphene can protect the topography of underlying DNA nanostructures against thermal deformation. For single-stranded DNAs (ss-DNAs), it can be attached on the graphene through π-π stacking interaction between the nucleobases and the graphene surface.[33-37] In comparison to the ss-DNAs, the DNA nanostructures used in this study do not have such an interaction with graphene, and only very weak van der Waals interactions can exist. This suggests that the transferred graphene is physisorbed on the DNA nanostructures. In comparison to the interactions between graphene and DNA nanostructures, interactions between graphene and a mica surface are stronger because surface charges on a mica surface can induce charge transfer to graphene, resulting in ionic bonds between graphene and the mica surface.[38] This suggests that the topography replication of graphene may originate mostly from the interactions between graphene and the mica surface, although interactions between graphene and the DNA nanostructures are not negligible. For this reason, the deformations of the DNA nanostructures themselves at high temperatures may be limited by the interactions between graphene and the mica surface because graphene needs to be locally detached from the mica surface for DNA nanostructures to be deformed. Therefore, the high thermal stability of the topography of the replicated graphene-DNA nanostructures may also originate mostly from the interactions between graphene and the mica surface. The enhanced thermal stability of the topography of the DNA nanostructures by the replicated graphene will widen the applications of the topography of the DNA nanostructures to devices fabricated through a thermal treatment process or processes operating at high temperatures.

The AFM experiments suggest that the 3D topography of the replicated graphene-DNA nanostructures can be changed with temperature. The 3D topography of graphene can influence its electronic structure, resulting in change of electrical properties.[12-14] For this reason, the sheet resistance of the replicated graphene-DNA nanostructures was measured using a four-point probe with increasing temperature to understand the relationship between the 3D topography and the electrical properties of the replicated graphene. We take the replicated graphene-5HRs rather than graphene-DX lattices because of the higher corrugated topography of the replicated graphene-5HRs. The sheet resistance of the graphene and the replicated graphene-5HRs was measured with increasing temperature at intervals of 5 °C, where the sheet resistance was recorded after waiting for 2 minutes at each temperature. The sheet resistance of pristine graphene without DNA nanostructures on the mica increased gradually with raising temperature (Fig. 3(b)). In comparison to the pristine graphene, the sheet resistance of replicated graphene-5HRs was different from pristine graphene (see the solid line in Fig. 3(c)). At temperatures below 70 °C, the sheet resistance gradually increased with raising temperature, resulting in a typical positive temperature coefficient that can be observed from pristine graphene.[37-39] Interestingly, when the temperature was above 70 °C, a negative temperature coefficient that cannot be found in pristine graphene was observed; the sheet resistance decreased with raising temperature. The negative temperature coefficient persisted up to approximately 130 °C. When temperature was raised above 130 °C, the sheet resistance again showed a positive temperature coefficient. Because both the 5HRs and mica surface are insulator, the changes in the sheet resistance can be ascribed to the electrical properties of the replicated graphene. The corrugated topography of graphene-5HRs is similar to a ripple of graphene. Highly corrugated graphene ripples were theoretically suggested to act as a long-range scattering potential for charge carriers, resulting in an increase in its resistance.[40] Thus, the increase in resistance of graphene-5HRs may have the same origin as that of the highly corrugated graphene ripples. These descriptions suggest that the negative temperature coefficient is related to the 3D topography of the replicated graphene. As observed in the AFM images in Fig. 2(d)-(f), the replicated graphene-5HRs were completely disassembled at a temperature of approximately 130 °C. When the 5HRs were completely disassembled, the graphene covers disassembled structures that were distributed randomly, as drawn schematically in Fig. 3(a). The temperature range of decreasing the sheet resistance is similar with that of the 5HRs disassembling. After the 5HRs were disassembled at a high temperature, the graphene-disassembled structures were cooled down to room temperature. The sheet resistance of the graphene-disassembled structures was lower than that of the replicated graphene-5HRs at room temperature. The lower sheet resistance at room temperature can be ascribed to the flattened topography of graphene on the disassembled structures. Subsequently, the sheet resistance of the graphene-disassembled structures was measured to compare it with the replicated graphene-5HRs (see the dashed line in Fig. 3(c)). The sheet resistance of the graphene-disassembled structures showed only a positive temperature coefficient with raising temperature, which is similar to that of the pristine graphene. The different behavior of the sheet resistance of the replicated graphene-DNA nanostructures from those of the pristine graphene and the graphene-disassembled structures suggests that the electric property of the replicated graphene is related to its 3D topography. The observed negative temperature coefficient of sheet resistance and the comparative experiments lead to the conclusion that the electrical property of graphene can be controlled by the 3D topography of the replicated graphene from designed DNA nanostructures.

In summary, the 3D topography of graphene was controlled by replication from designed DNA nanostructures, such as 5HRs and DX lattices. Owing to the variety of DNA nanostructures, graphene topography with particular shapes and sizes can be designed. Furthermore, because this replication was achieved using CVD graphene, it can be scaled up when DNA nanostructures can be distributed appropriately. The enhanced thermal stability of the topography of graphene-covered DNA nanostructures provides the opportunity for the topography of DNA nanostructures to be applied to devices and sensors operating at high temperatures. The observed negative temperature coefficient of resistance of the replicated graphene suggests that the electrical properties of graphene can be controlled by the replication from designed DNA nanostructures, which opens a possibility for applications of DNA nanostructures to temperature-dependent electrical fuses or sensors for channel breaking.

This work was supported by a National Research Foundation of Korea (NRF) grant funded by the Korean government (MEST) (No. 2012R1A1A2041241) and Basic Science Research Program through the National Research Foundation of Korea (NRF) funded by the Ministry of Education (2009-0094023).

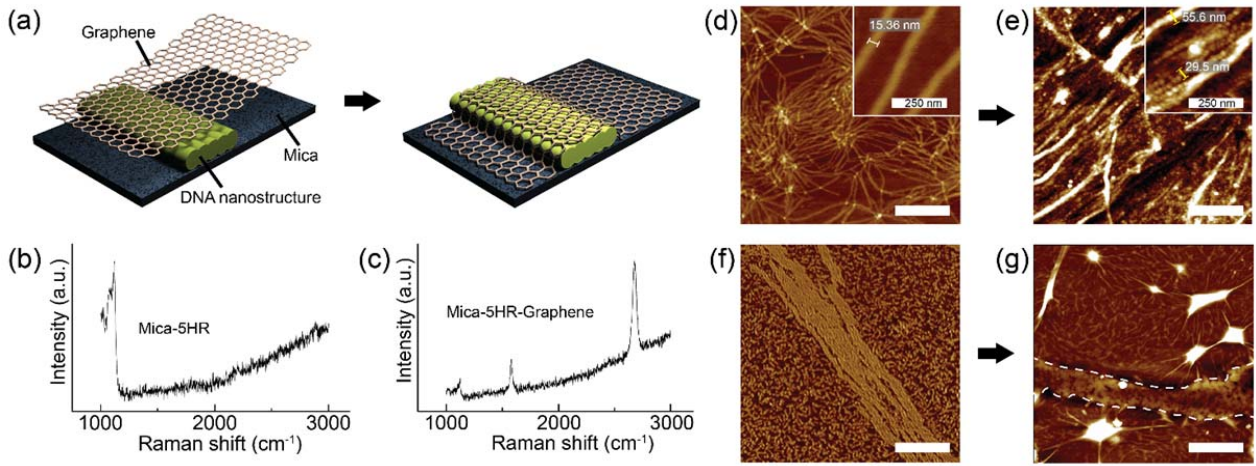

FIG. 1. (Color online) (a) Schematic drawings for the replication of graphene by DNA nanostructures. The light green cylinders on a mica surface in (a) indicate the DNA duplexes and these 5 duplexes form a single 5HR. The mesh on the 5HR in (a) represents the transferred CVD graphene onto the 5HR on the mica surface. (b)-(c) Raman spectra of the 5HRs on a mica surface (b) before and (c) after the transfer of graphene. (d)-(e) AFM images of (d) 5HRs and (e) replicated graphene-5HRs. (f)-(g) AFM images of (f) DX lattices and (g) replicated graphene-DX lattices. The white dashed lines in (g) denote replicated graphene-DX lattices. All scale bars in (d)-(g) are 500 nm.

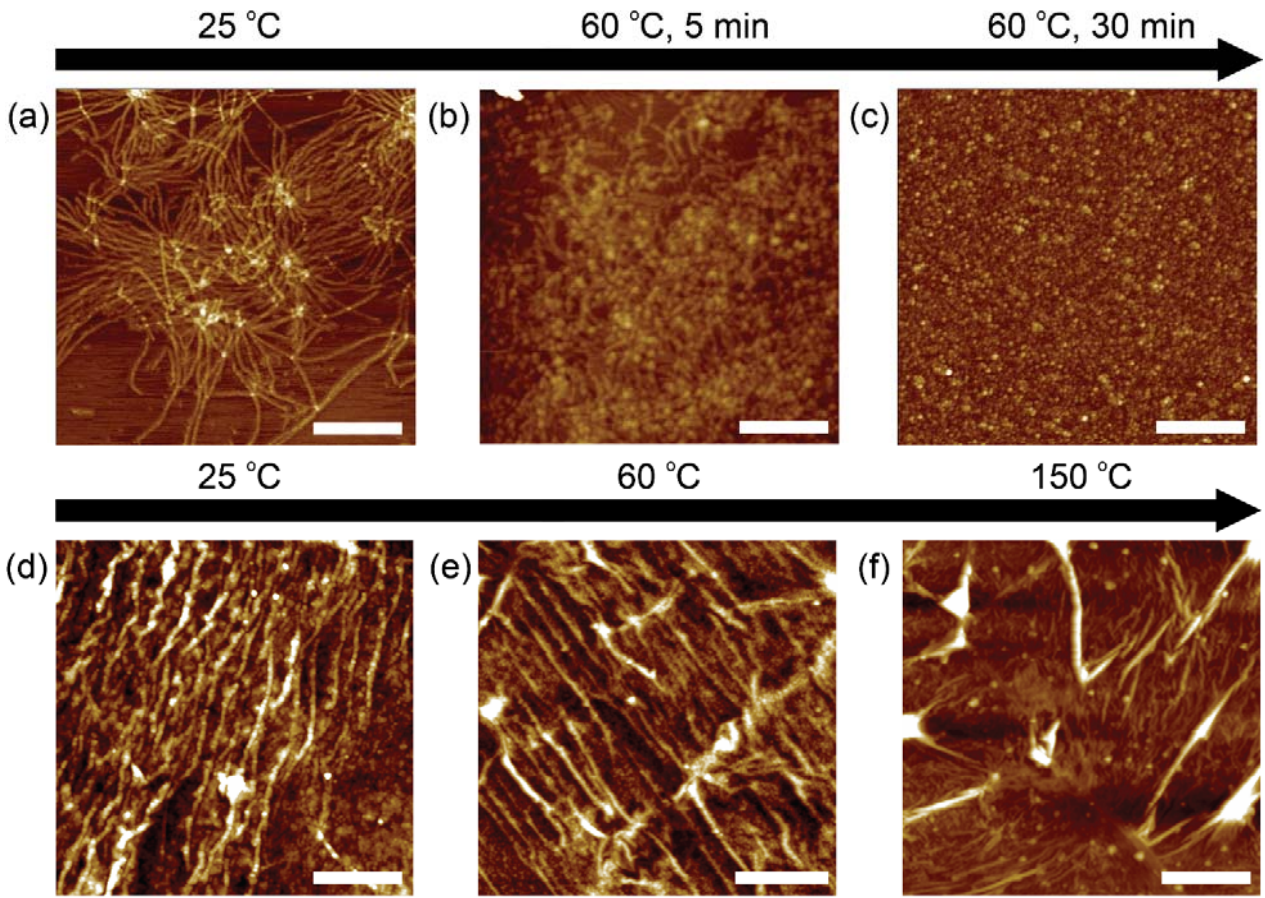

FIG. 2. (Color online) (a)-(c) AFM images of 5HRs on a mica surface measured after heating at different temperatures: (a) 25 °C, (b) 60 °C for 5 minutes, (c) 60 °C for 30 minutes. (d)-(f) AFM images of replicated graphene-5HRs on a mica surface measured after heating at different temperatures: (d) 25 °C, (e) 60 °C for 30 minutes, (f) 150 °C for 30 minutes. All scale bars in the AFM images are 250 nm.

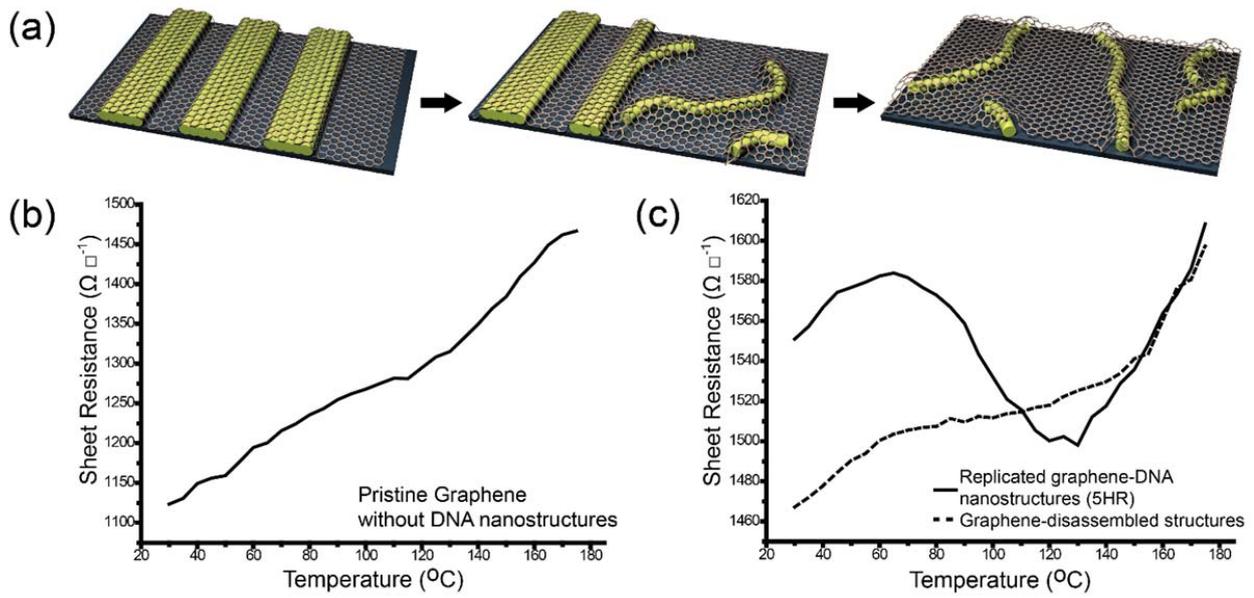

FIG. 3. (Color online) (a) Schematic drawings for topographical changes in replicated graphene-DNA nanostructures (5HR) with increasing temperature. The 5HRs are denatured and deformed at high temperatures. (b) Changes in the sheet resistance of pristine graphene without DNA nanostructures with raising temperature. (c) Changes in the sheet resistance of replicated graphene-DNA nanostructures (5HR) and graphene-disassembled structures with raising temperature, where the solid and dashed lines denote sheet resistance of graphene-DNA nanostructures and graphene-disassembled structures, respectively.